\documentclass{article}

\usepackage{arxiv}

\usepackage[utf8]{inputenc} 
\usepackage[T1]{fontenc}    
\usepackage{hyperref}     
\usepackage{algorithmic}
\usepackage{url}   
\usepackage{fancyhdr}
\usepackage{booktabs}    
\usepackage{caption}
\usepackage{subcaption}
\usepackage{amsfonts} 
\usepackage{textcomp}
\usepackage{nicefrac}       
\usepackage{microtype} 
\usepackage{balance}
\usepackage{xcolor}
\usepackage{verbatim}
\usepackage{lipsum}
\usepackage{graphicx}
\graphicspath{ {./images/} }

\title{Geospatial and Temporal Trends in Urban Transportation: A Study of NYC Taxis and Pathao Food Deliveries}

\author{
 Bidyarthi Paul \\
  Department of Computer Science and Engineering\\
  Ahsanullah University of Science and Technology\\
  Dhaka, Bangladesh \\
  \texttt{bidyarthi01@gmail.com} \\
   \And
Fariha Tasnim Chowdhury \\
 Department of Computer Science and Engineering\\
  Ahsanullah University of Science and Technology\\
  Dhaka, Bangladesh \\
  \texttt{farihatasnim01@gmail.com} \\
  \And
  Dipta Biswas \\
  Department of Computer Science and Engineering\\
  Ahsanullah University of Science and Technology\\
  Dhaka, Bangladesh \\
  \texttt{diptabiswas13@gmail.com} \\
\And
   Meherin Sultana \\
 Department of Computer Science and Engineering\\
  Ahsanullah University of Science and Technology\\
  Dhaka, Bangladesh \\
  \texttt{meherinsultana36@gmail.com} \\
}

\begin{document}
\maketitle
\begin{abstract}
Urban transportation plays a vital role in modern city life, affecting how efficiently people and goods move around. This study analyzes transportation patterns using two datasets: the NYC Taxi Trip dataset from New York City and the Pathao Food Trip dataset from Dhaka, Bangladesh. Our goal is to identify key trends in demand, peak times, and important geographical hotspots. We start with Exploratory Data Analysis (EDA) to understand the basic characteristics of the datasets. Next, we perform geospatial analysis to map out high-demand and low-demand regions. We use the SARIMAX model for time series analysis to forecast demand patterns, capturing seasonal and weekly variations. Lastly, we apply clustering techniques to identify significant areas of high and low demand. Our findings provide valuable insights for optimizing fleet management and resource allocation in both passenger transport and food delivery services. These insights can help improve service efficiency, better meet customer needs, and enhance urban transportation systems in diverse urban environments.
\end{abstract}


\section{Introduction}
Modern city life relies significantly on urban transportation, which affects how efficiently people and goods can move around. It includes a range of transportation options that are essential to cities' everyday operations, such as delivery services, public transportation, and taxis. Strong urban transportation systems are necessary to alleviate congestion, reduce travel times, and improve accessibility. This will help agencies and operators significantly improve service all around by providing a clear look into these systems' patterns and needs. Analyzing urban transportation is necessary because it helps city planners and service providers identify peak demand periods, optimize fleet management, and allocate resources more effectively. With growing urban populations and increasing demand for timely services, it becomes crucial to enhance the efficiency of transportation networks.

In our study, we aim to analyze urban transportation patterns using two distinct datasets. Our objective is to identify key trends in demand, peak times, and important geographical hotspots. By examining these datasets, we seek to provide actionable insights for optimizing fleet management and resource allocation in both passenger transport and food delivery services.

We are using two datasets from different cities—the NYC Taxi Trip Dataset from New York City and the Pathao Food Trip Dataset from Dhaka, Bangladesh—to gain a comprehensive understanding of urban transportation patterns. Although these datasets represent different types of services—passenger transport and food delivery—they both capture essential aspects of urban mobility and demand. By analyzing data from these two distinct urban environments, we can identify common trends such as peak demand periods, geographical hotspots, and service efficiency. This comparative approach allows us to draw broader conclusions about urban transportation.

\begin{figure}[h]
    \centering
    \includegraphics[width=0.8\textwidth]{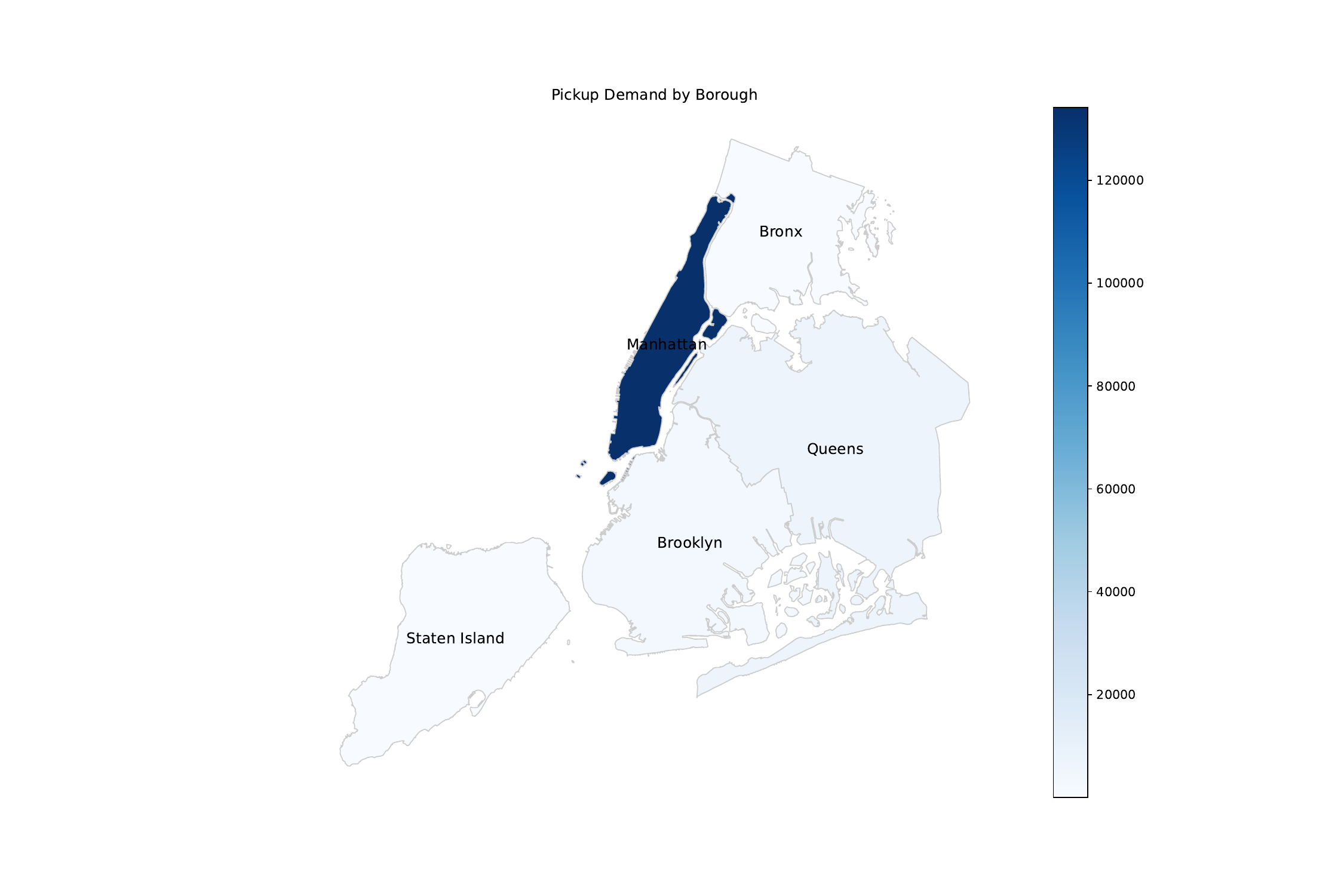}
    \caption{Geospatial area in New York City}
    \label{fig:study}
\end{figure}

In this study, we begin with Exploratory Data Analysis (EDA) to gain a basic understanding of the datasets and identify key trends. EDA involves visualizing and summarizing the data to uncover patterns, relationships, and anomalies. This helps us understand how taxi and food delivery services are used over time and across different locations. Next, we perform geospatial analysis to examine the distribution of trips across various areas. This analysis allows us to map out high-demand and low-demand regions, showing where most pickups occur. By visualizing these patterns, we can see which areas are busiest and when they experience the most activity. For time series analysis, we use the SARIMAX model to forecast demand patterns. This model is particularly useful because it can handle both seasonal and non-seasonal data, capturing regular fluctuations over time. By analyzing the data with SARIMAX, we can predict future demand for taxis and food deliveries, helping service providers prepare for busy periods and optimize their operations. Lastly, we apply clustering techniques to identify significant regions of high and low demand. K-means clustering groups similar geographic locations based on the frequency of trips. This method helps us pinpoint specific neighborhoods or areas that consistently have high or low demand, providing insights into where resources should be allocated. 

Our objective is to uncover valuable insights that can help improve fleet management and service delivery in urban environments. These insights will enable taxi and food delivery services to better meet customer needs, enhance efficiency, and optimize resource allocation. Through our analysis, we hope to contribute to more effective urban transportation systems.

\section{Related Works}

In a study of Archie et al. (2023)~\cite{3}, the New York Taxi Fare Prediction dataset (2009-2015) is examined to derive insights into the dynamics of urban transportation. This employs mean imputation techniques and spatial visualization. The analysis reveals significant patterns in taxi demand, particularly within the boroughs of Brooklyn and Manhattan. These findings underscore the critical role of data in optimizing urban mobility and informing transportation planning strategies.

Some advancements in Geographic Information Systems (GIS) and spatial analysis have highlighted their transformative potential across various domains, including educational planning, urban mobility, and climate change impact assessment. For instance, Kadir et al. (2016)~\cite{1} utilized GIS and Inverse Distance Weighting (IDW) interpolation to analyze and map trends in Science and Mathematics GPA among schools in Kelantan, Malaysia, from 2010 to 2014. This analysis demonstrates GIS's efficacy in enhancing educational planning and decision-making processes. In a related domain, Xie, Chen, et al. (2021)~\cite{2} leveraged public datasets from the New York City Taxi and Limousine Commission (TLC) to explore traffic demand and community structures using spatial statistics. This research revealed power-law distributions within taxi zones and provided insights into community interactions through graph-based approaches. The impact of climate change on the Kolkata Metropolitan Area over 35 years was investigated by Talapatra et al. (2021)~\cite{10} using satellite image analysis, where various indices such as Land Surface Temperature (LST), Normalized Difference Vegetation Index (NDVI), and Normalized Difference Built-up Index (NDBI) were calculated through advanced GIS techniques and statistical methods, underscoring GIS's crucial role in environmental monitoring and climate change assessment.

There have been some developments in approaches to time series forecasting across various fields. Khan et al. (2020)~\cite{4} evaluates ARIMA models for predicting Netflix stock prices using data from 2015 to 2020. Among the models tested, ARIMA (1, 1, 33) proved to be the most accurate. Another study by Albeladi et al. (2023)~\cite{5} compares LSTM and ARIMA models, showing that ARIMA performed better with lower errors. Both models serve different forecasting needs, with ARIMA excelling in this case. Other research like Sirisha et al. (2022)~\cite{6} focuses on forecasting profits using ARIMA, SARIMA, and LSTM models. ARIMA and SARIMA showed a gradual decline in profits, while LSTM reflected a more sudden decrease. Another study by Astuti et al. (2018)~\cite{7} focuses on forecasting train passengers between Surabaya and Jakarta and found that the SARIMA (0,1,1)(1,1,0)12 model was the most accurate. These studies show that while ARIMA models are reliable for forecasting, LSTM also has potential, especially for capturing complex patterns.

There are studies focused on improving clustering methods to enhance data analysis. Sinaga et al. (2020)~\cite{9} introduces an advanced algorithm called Unsupervised k-means (U-k-means). This algorithm improves traditional k-means by adding an entropy penalty term, which helps it find the best number of clusters and reduces sensitivity to starting points. The method uses flexible initialization and iteratively refines clusters. Tests on various datasets show that U-k-means is more accurate, robust, and adaptable than traditional k-means.

\section{Dataset}

We used the NYC Taxi Trip Dataset and Pathao Food Trip to perform our comparative analysis. Together, this gives us the ability to study geospatial and temporal patterns in taxi services and food delivery trips. Our study primarily focuses on the NYC Taxi Trip Data, which provides comprehensive information on taxi trips in New York City. Since a suitable taxi trip dataset for Bangladesh is unavailable, we included the Pathao Food Trip dataset as a substitute. This dataset captures temporal characteristics related to food delivery trips. While the NYC Taxi Trip Data is the main subject of our analysis, the Pathao Food Data will augment our research by offering additional perspectives on transportation and delivery trends in Dhaka, Bangladesh.

\subsection{Data Collection}

\subsubsection{NYC Taxi Trip Dataset}
The ``NYC Taxi Trip Data" dataset was obtained from the New York City Taxi and Limousine Commission\footnote{\url{https://www.andresmh.com/nyctaxitrips/}}. The study area of our dataset where most trips have been made is in Figure~\ref{fig:study}. It focuses on taxi trips of January 2013 and totals approximately 2.5 GB before decompression. Although the dataset spans the entire year of 2013, we concentrated on January due to the data's large size. By focusing on a single month, we ensure quicker model construction, faster code execution, and improved iterations and changes. A thorough picture of taxi usage and regional mobility patterns in New York City is provided by each entry, which includes specifics like the pickup datetime, dropoff datetime, passenger count, trip time in secs, trip distance, pickup longitude, pickup latitude, dropoff longitude, and dropoff latitude.

\subsubsection{Pathao Food Trip Dataset}

The "Pathao Food Trip Dataset" is extracted from a GitHub repository\footnote{\url{https://github.com/AwsafAlam/Pathao_Food_Recommendation}} that is available to the public. The dataset consists of 323,139 records, which include data on user IDs, item IDs, day of the week, hour of the day, category IDs, restaurant IDs, cuisine IDs, and item counts. While this dataset offers an accurate representation of trends in food delivery services, it does not contain any geospatial coordinates. So, geospatial analysis can not be done here. Additionally, it lacks datetime attributes, which prevents us from performing time series analysis. However, we will conduct Exploratory Data Analysis (EDA) on this dataset to uncover key trends and insights.

\subsection{Data Pre-processing}
For the ``NYC Taxi Trip Dataset" and ``Pathao Food Trip Dataset," we cleaned the data by handling missing values and removing duplicates. By utilizing feature engineering techniques, new features were created from the existing features. Specifically, we extracted the hour, day, and month from the pickup datetime, as well as the day of the week, to capture various time-related aspects of each trip. Additionally, we computed the trip duration in minutes by calculating the difference between dropoff datetime and pickup datetime. These new features provide deeper insights into temporal patterns and trip characteristics, facilitating more detailed analysis.

\subsection{Train-Test Splits}
To evaluate the performance of our models, we performed an 80:20 train-test split on both datasets.

\section{Methodology}
\subsection{Exploratory Data Analysis (EDA)}

\subsubsection{NYC Taxi Trip Dataset}
\begin{figure*}[ht]
    \centering
    \begin{subfigure}[b]{0.32\textwidth}
        \centering
        \includegraphics[width=\textwidth]{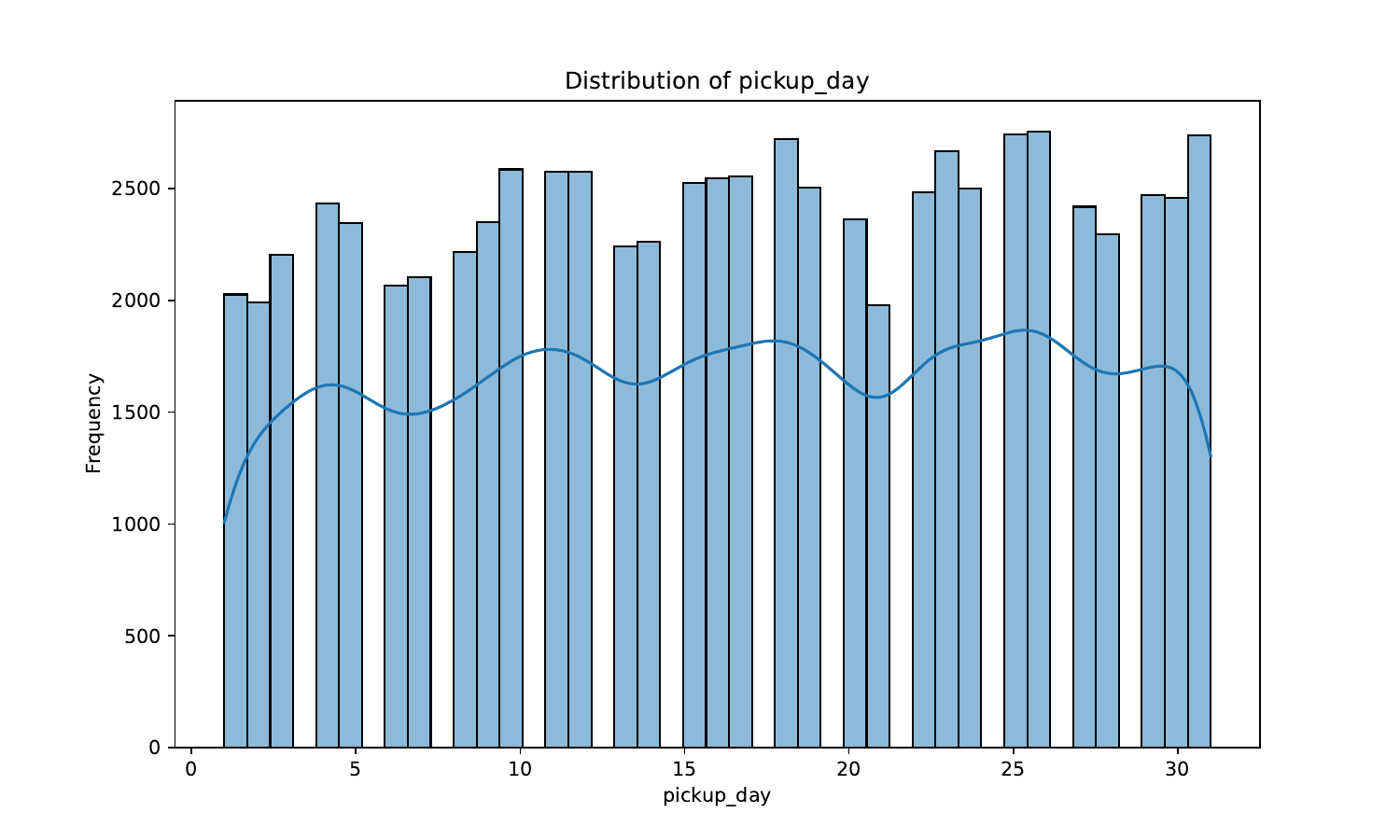}
        \caption{Distribution of Pickup day}
        \label{fig:image1}
    \end{subfigure}
    \hfill
    \begin{subfigure}[b]{0.32\textwidth}
        \centering
        \includegraphics[width=\textwidth]{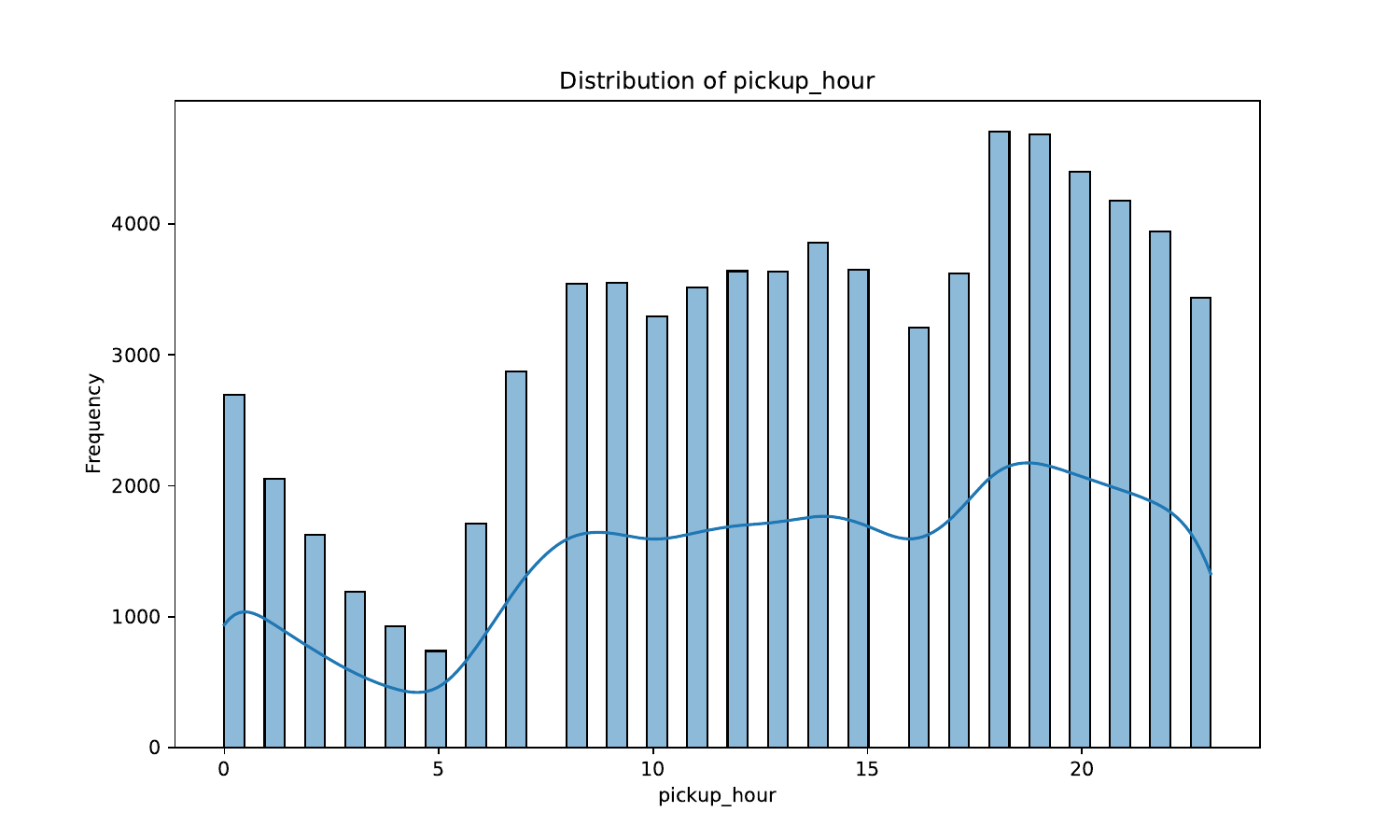}
        \caption{Distribution of Pickup hour}
        \label{fig:image2}
    \end{subfigure}
    \hfill
    \begin{subfigure}[b]{0.32\textwidth}
        \centering
        \includegraphics[width=\textwidth]{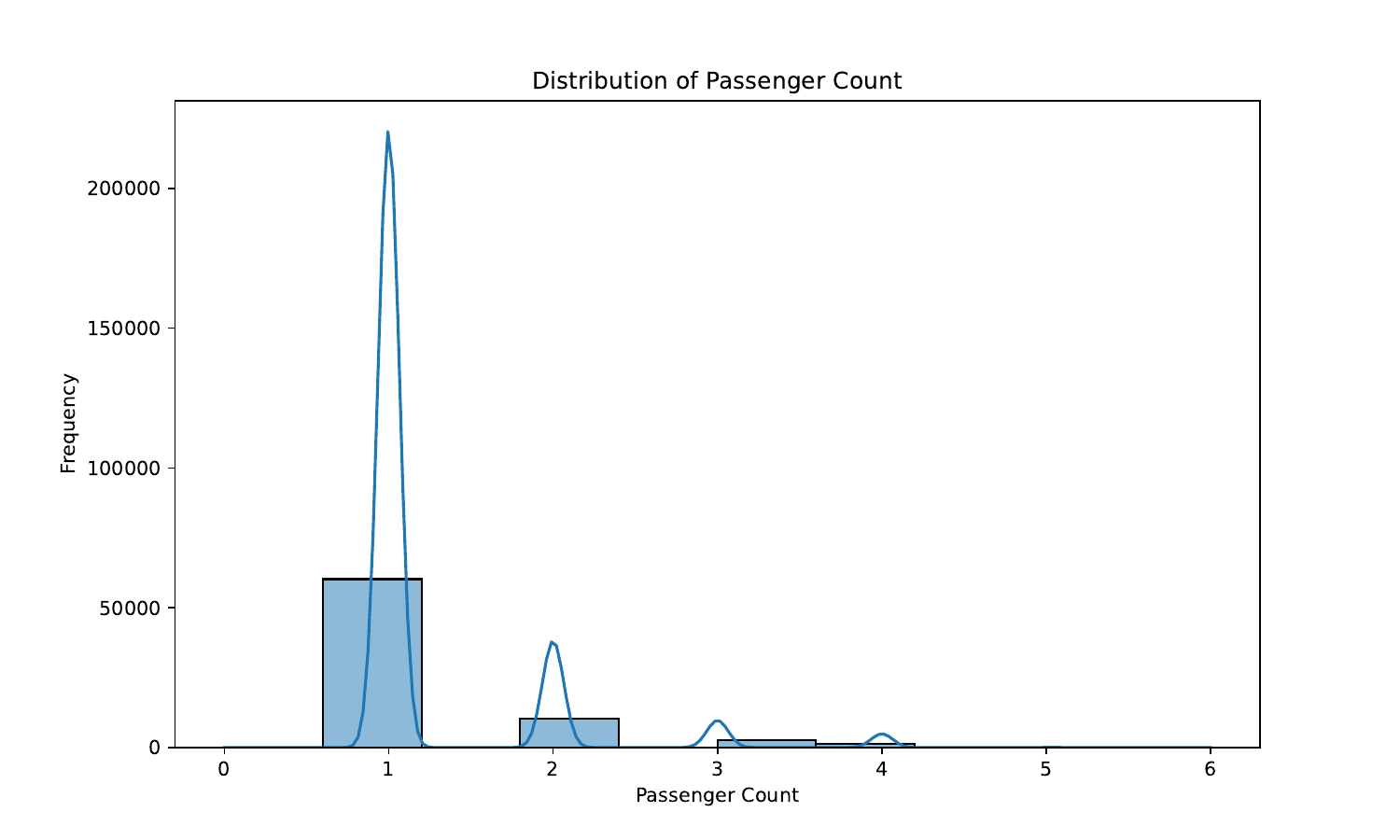}
        \caption{Frequency of Passenger Count}
        \label{fig:image3}
    \end{subfigure}
    \hfill
    \begin{subfigure}[b]{1.0\textwidth}
        \centering
        \includegraphics[width=\textwidth]{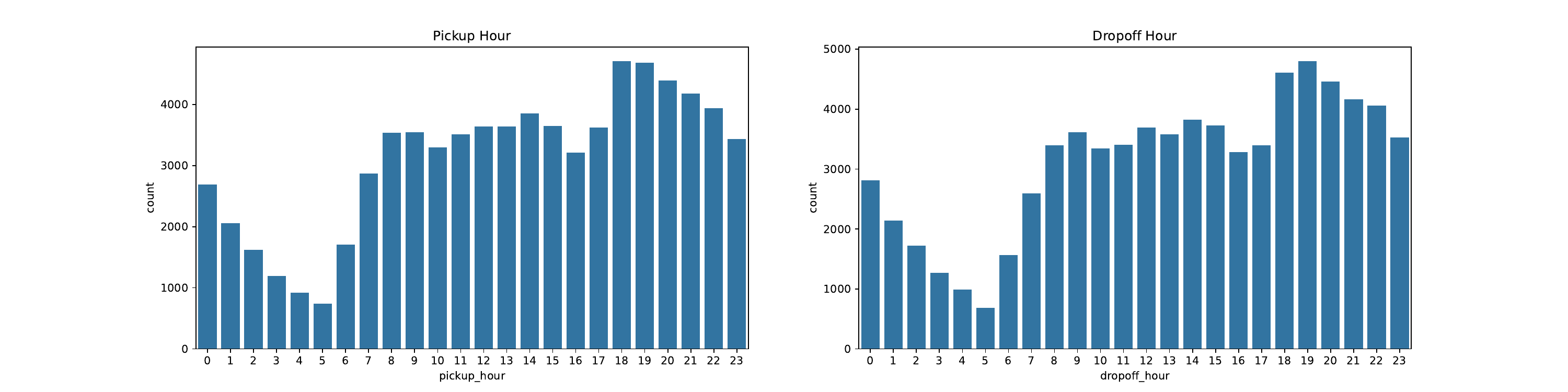}
        \caption{Pickup and Dropoff time in the hour}
        \label{fig:image4}
    \end{subfigure}
    \hfill
    \begin{subfigure}[b]{0.45\textwidth}
        \centering
        \includegraphics[width=\textwidth]{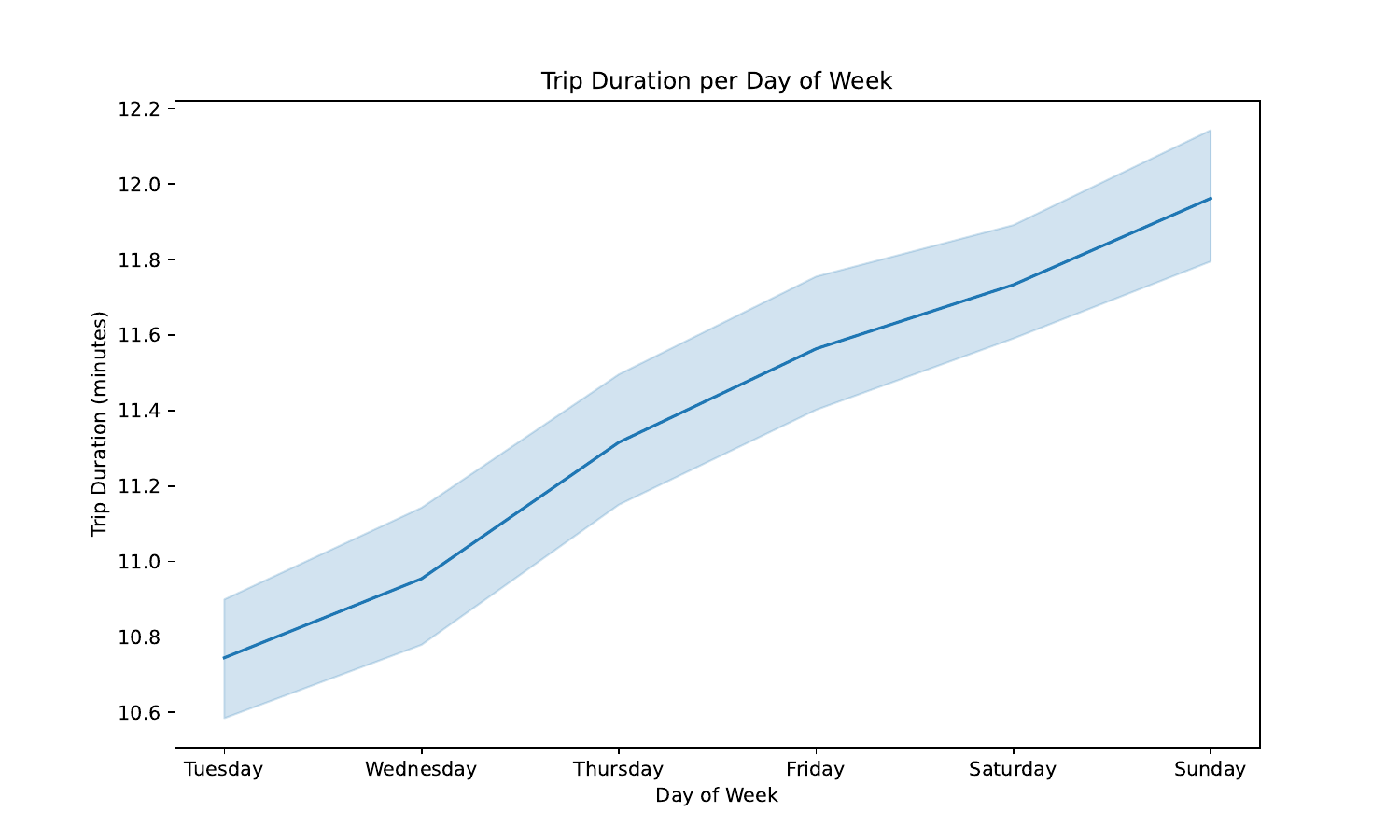}
        \caption{Trip Duration per Day of Week}
        \label{fig:image5}
    \end{subfigure}
    \hfill
    \begin{subfigure}[b]{0.45\textwidth}
        \centering
        \includegraphics[width=\textwidth]{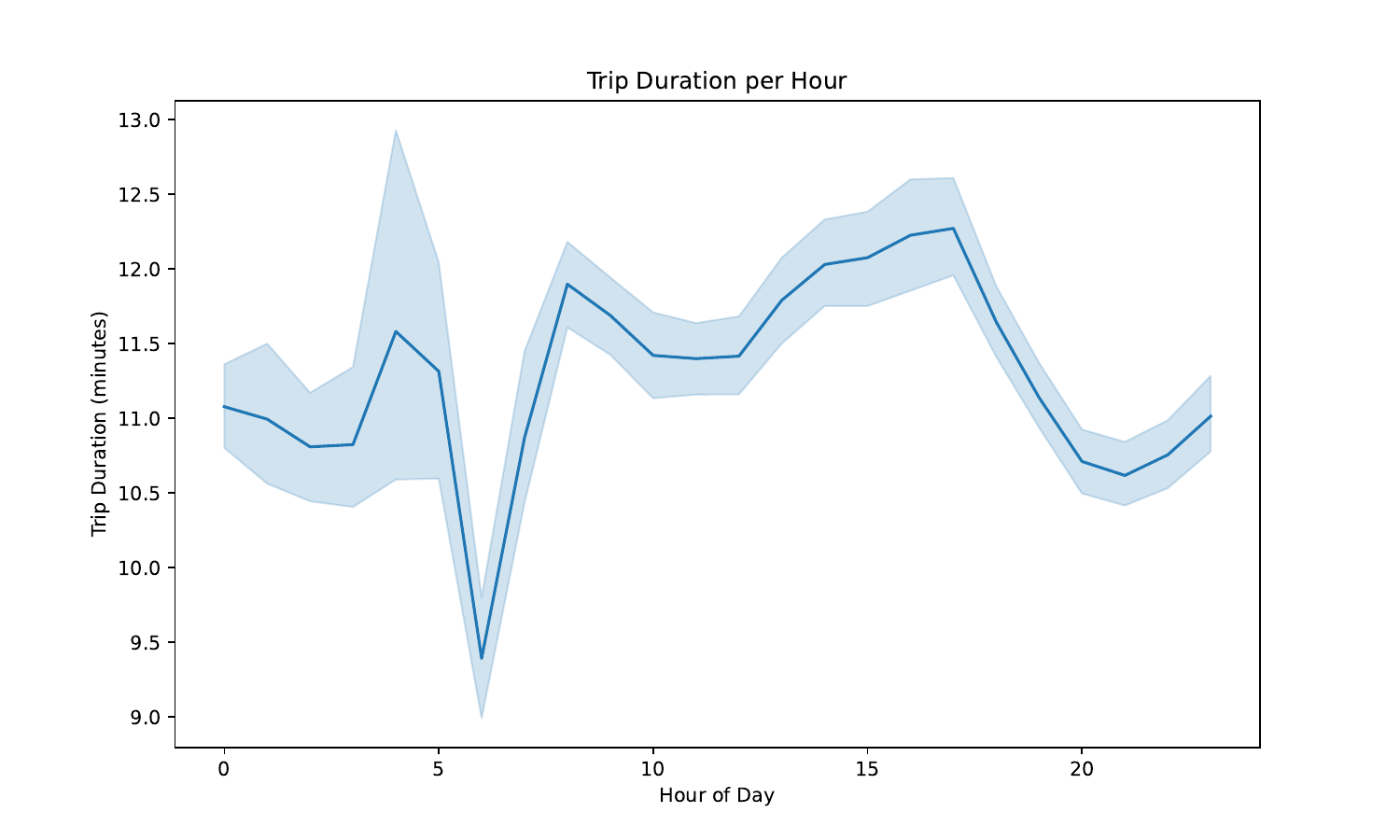}
        \caption{Trip Duration per Hour of Day}
        \label{fig:image6}
    \end{subfigure}
    \caption{EDA on NYC Taxi Trip Dataset}
    \label{fig:three}
\end{figure*}

To gain a comprehensive understanding of the NYC Taxi dataset, we performed a series of exploratory data analyses focusing on various temporal and trip characteristics that can be seen in Figure~\ref{fig:three}. The analyses included:

\begin{itemize}

    \item \textbf{Correlation Analysis:}
    We performed a correlation analysis between pickup day, pickup hour, trip duration, and trip distance to identify significant correlations, which can reveal patterns such as how the time of day or day of the week impacts trip duration and distance. It has been illustrated in Figure~\ref{fig:mat}.

    \begin{figure}[h]
    \centering
    \includegraphics[width=0.7\textwidth]{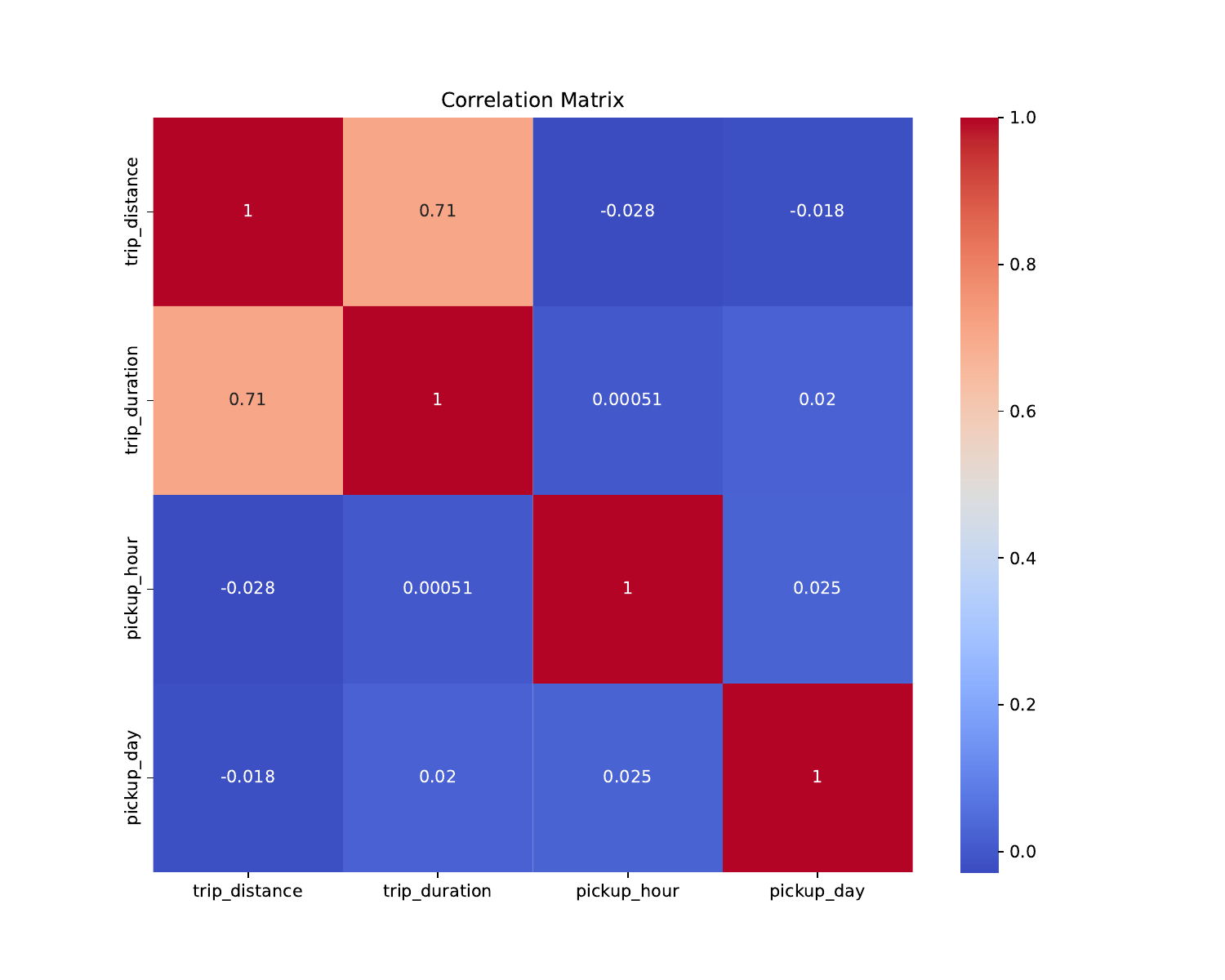}
    \caption{Correlation Matrix}
    \label{fig:mat}
    \end{figure}

    \item \textbf{Distribution of Pickup Hour:}
     Analyzed the distribution of taxi pickups by hour of the day to identify peak hours and periods of high demand.
    
    \item \textbf{Distribution of Pickup Day:}
    Examined the distribution of taxi pickups by day of the month to understand daily trends in taxi usage through the month. 
    
    \item \textbf{Distribution of Passenger Count:}
  Investigated the distribution of the number of passengers per trip to gain insights into the typical occupancy rates of taxis. 
    
    \item \textbf{Pickup and Dropoff Time of Hour:}
    Explored the pickup and dropoff times on an hourly basis to gain a finer granularity of the taxi activity throughout the day.
    
    \item \textbf{Trip Duration by Day of the Week and Hour of the Day:}
    Inspected the trip duration across different days of the week and hours of the day to identify patterns in trip lengths. 

\end{itemize}

\begin{figure}[ht]
    \centering
    \begin{subfigure}[b]{0.5\textwidth}
        \centering
        \includegraphics[width=\textwidth]{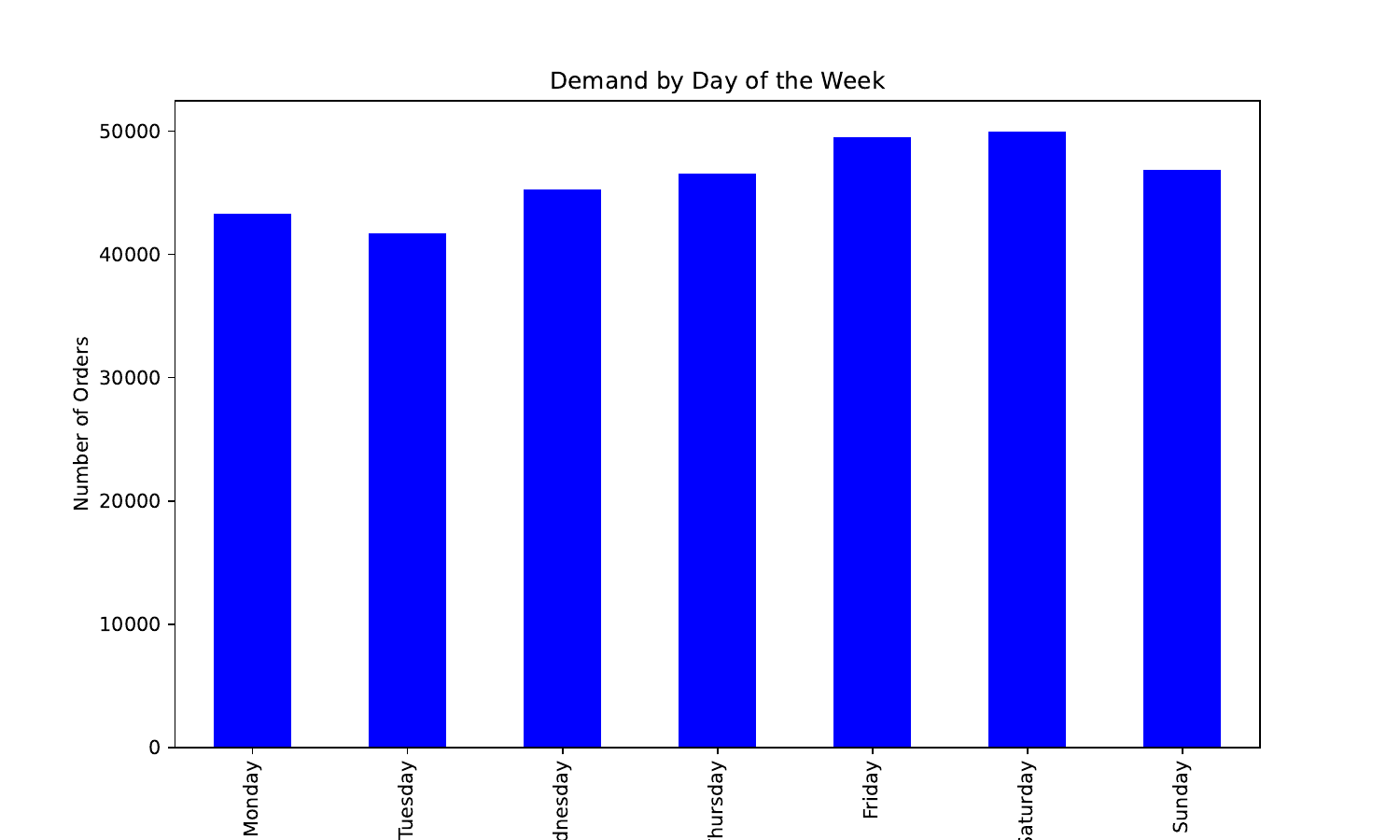}
        \caption{Demand By Day of the Week}
        \label{fig:im1}
    \end{subfigure}
    \hfill
    \begin{subfigure}[b]{0.5\textwidth}
        \centering
        \includegraphics[width=\textwidth]{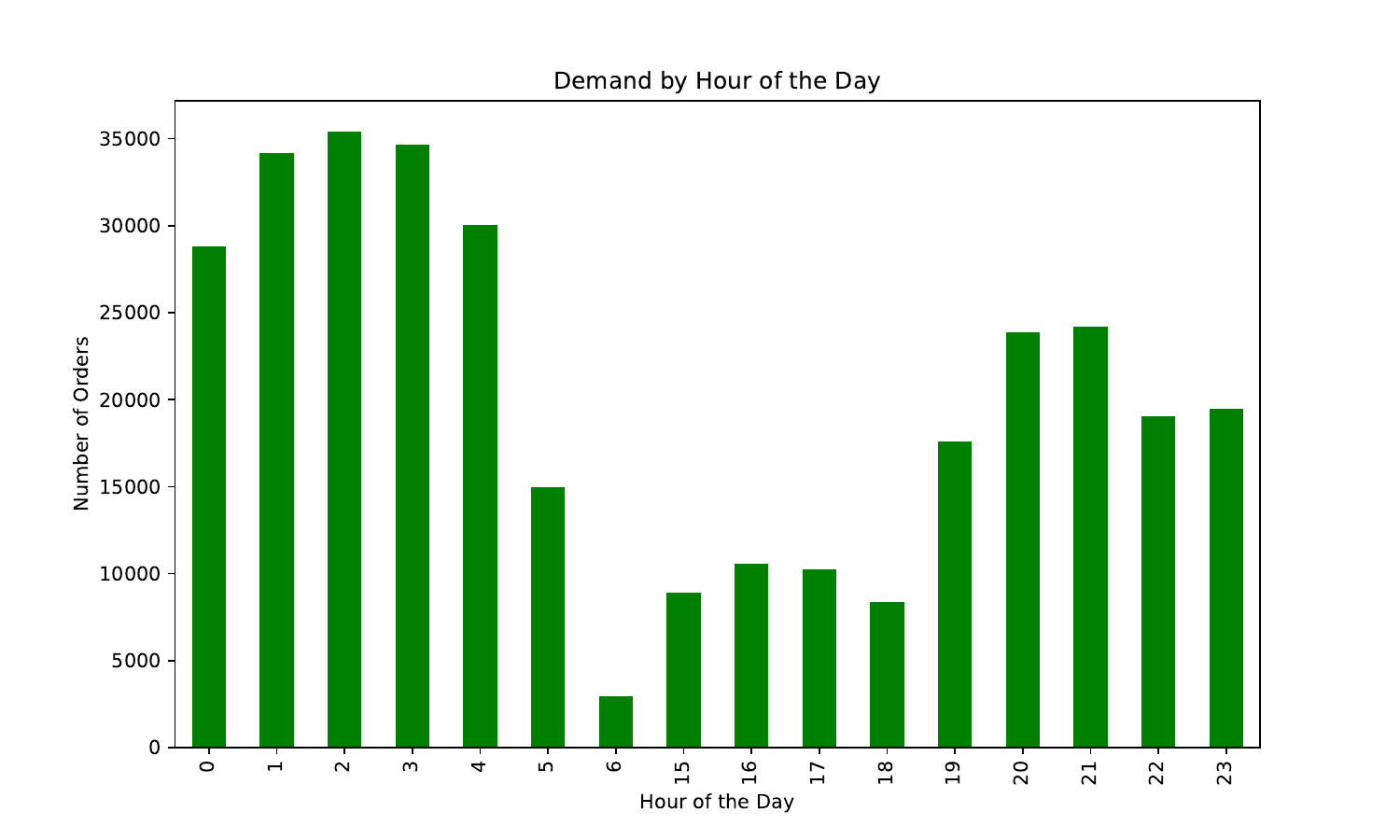}
        \caption{Demand By Hour of the Day}
        \label{fig:im2}
    \end{subfigure}
    \hfill
    \begin{subfigure}[b]{0.5\textwidth}
        \centering
        \includegraphics[width=\textwidth]{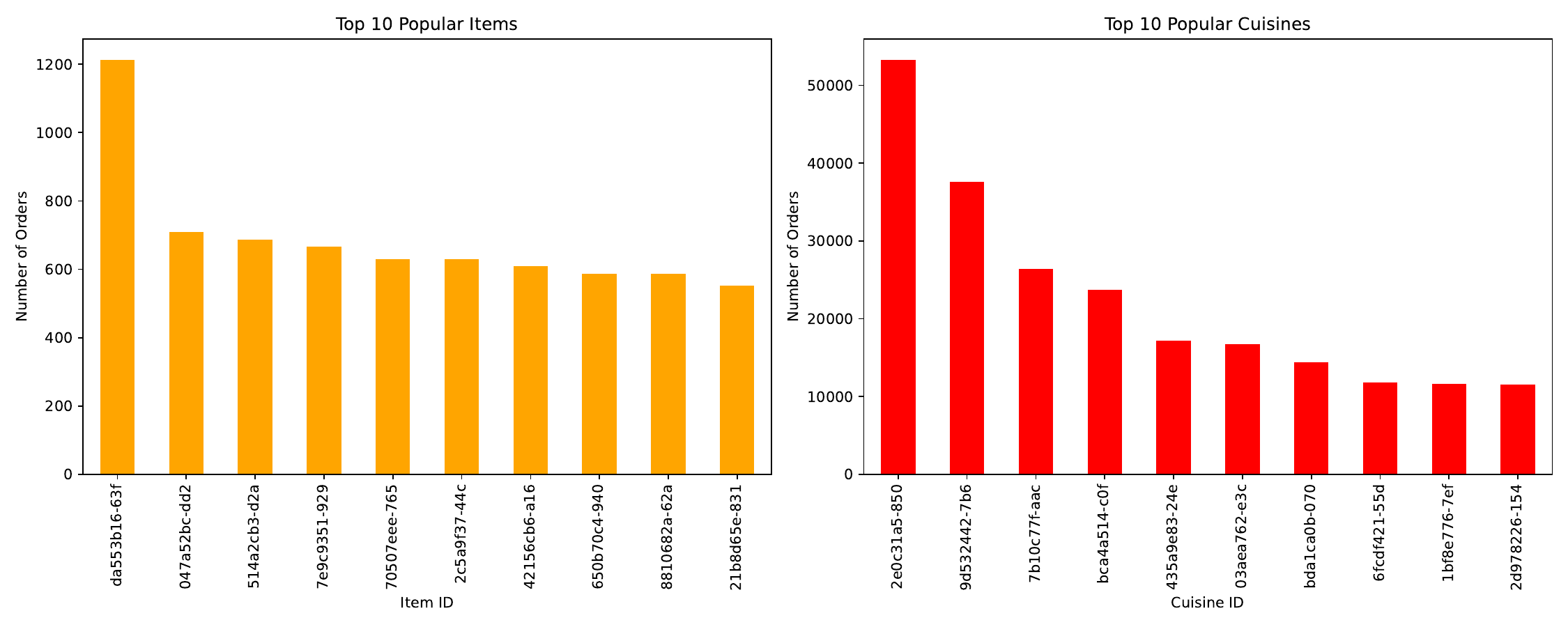}
        \caption{Top 10 Popular Items and Cuisines}
        \label{fig:ima3}
    \end{subfigure}
    \caption{EDA on Pathao Food Trip Dataset}
    \label{fig:three_images}
\end{figure}

\subsubsection{Pathao Food Trip Dataset}

To gain insights into the Pathao Food Trip Dataset, we performed the following analyses and their visual illustrations can be seen in Figure~\ref{fig:three_images}:

\begin{itemize}
    \item \textbf{Demand by Day of the Week (DoW)}: We analyzed the demand for food delivery services based on the day of the week to understand weekly demand patterns.
    
    \item \textbf{Demand by Hour of the Day (HoD)}: We examined the hourly demand for food deliveries to identify peak times for orders throughout the day. 
    
    \item \textbf{Top 10 Popular Items}: To identify the most frequently ordered items, we ranked the items based on their order counts. 
    
    \item \textbf{Top 10 Popular Cuisines}: Similarly, we ranked the cuisines based on the number of orders to determine the most popular cuisines. 
\end{itemize}

\subsection{Geospatial Analysis}

\begin{figure}[h]
    \centering
    \includegraphics[width=1.0\textwidth]{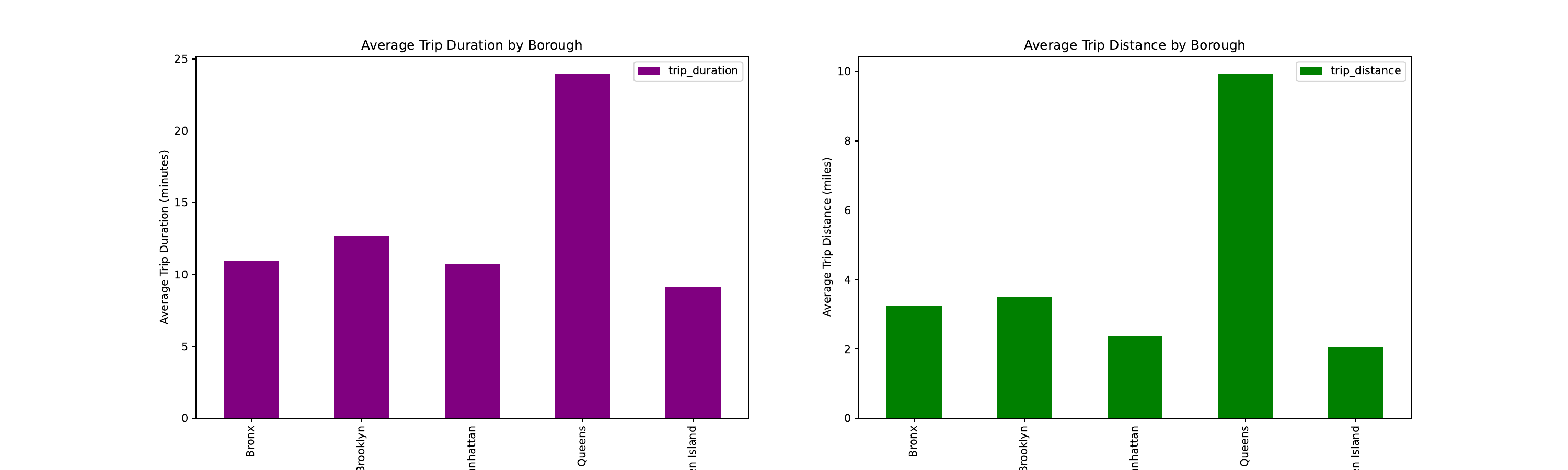}
    \caption{Average Trip Duration and Distance by Borough}
    \label{fig:avg}
\end{figure}

In the geospatial analysis, we examined the pickup demand in Figure~\ref{fig:study} across different boroughs of New York City (Bronx, Brooklyn, Manhattan, Queens, and Staten Island). We also analyzed the average trip distance and duration for each borough in Figure~\ref{fig:avg}. This analysis provides valuable insights into areas with varying levels of taxi demand. Understanding the average trip distances and durations by borough offers a detailed view of travel patterns.

\subsection{Time Series Analysis}

\begin{figure}[h]
    \centering
    \includegraphics[width=1.0\textwidth]{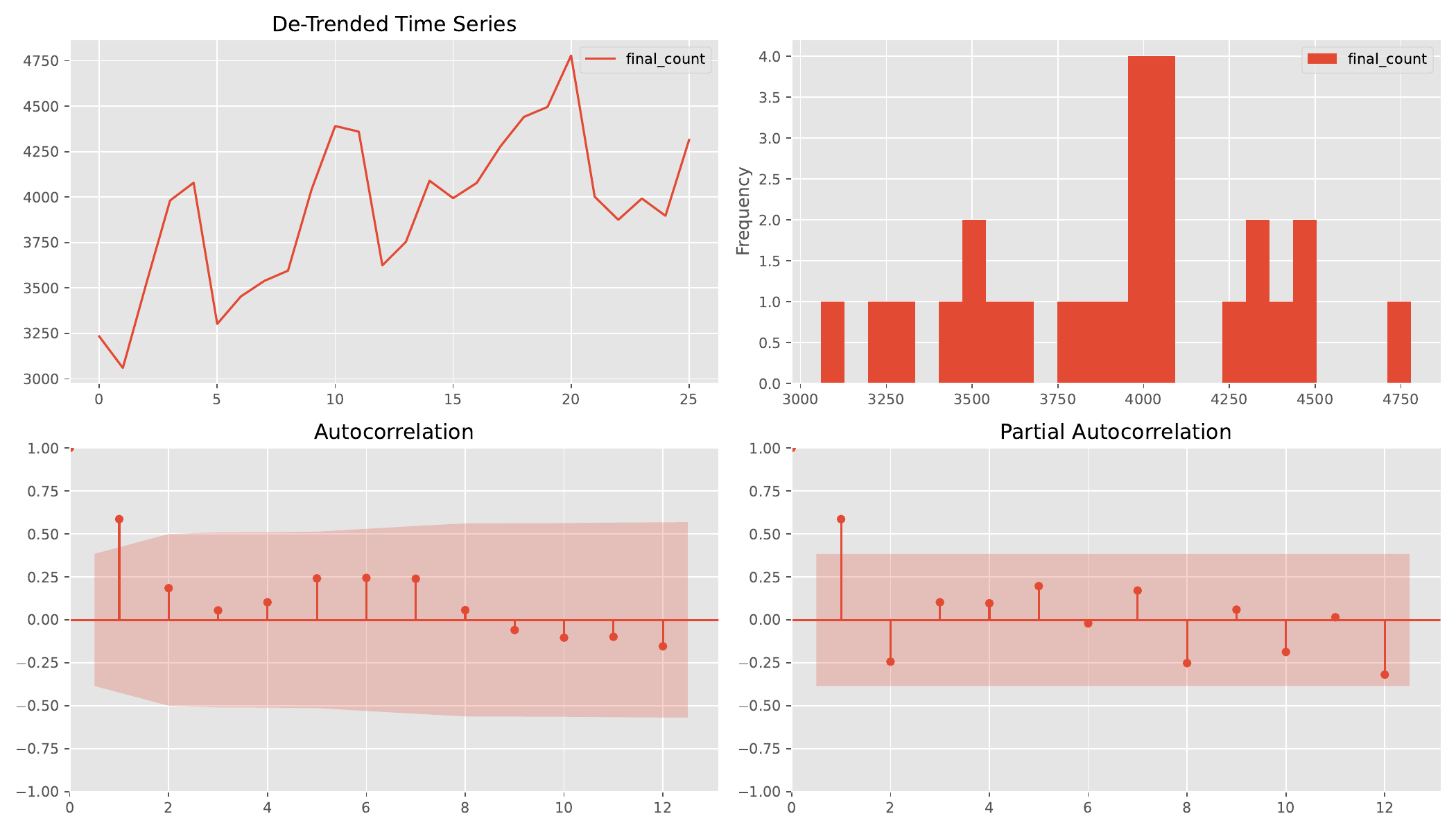}
    \caption{Detrending the Seasonal Data}
    \label{fig:pac}
\end{figure}

\begin{figure}[h]
    \centering
    \includegraphics[width=0.7\textwidth]{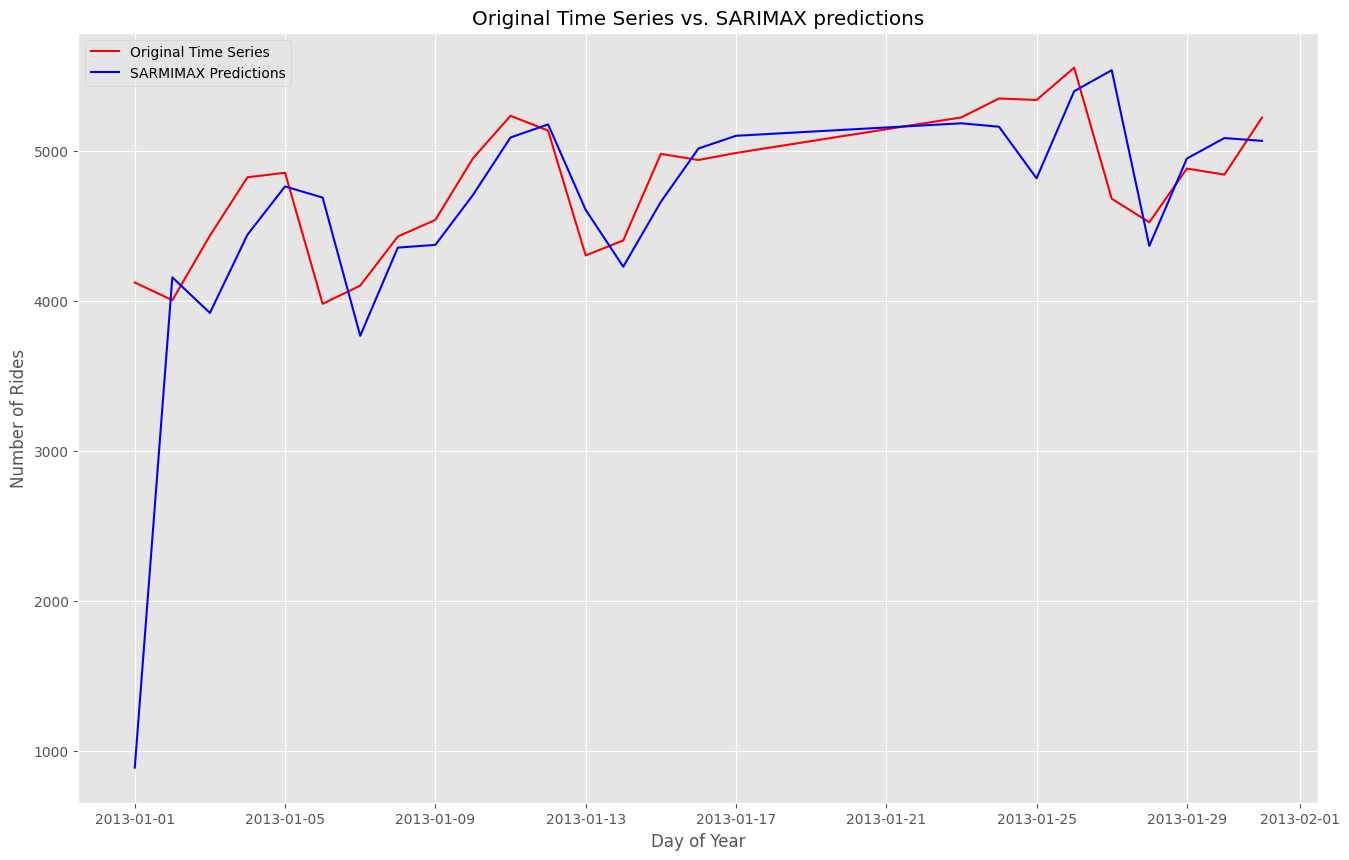}
    \caption{Predicting the number of rides using SARIMAX model}
    \label{fig:sari}
\end{figure}

In the time series analysis, we utilized a Seasonal Autoregressive Integrated Moving Average with Exogenous Variables (SARIMAX) model on the NYC Taxi Trip Data. The dataset displayed clear seasonal patterns, making a model capable of addressing seasonality essential. SARIMAX was particularly necessary as it can integrate both seasonal effects and exogenous variables, resulting in a more accurate and comprehensive prediction model. By de-seasonalizing the data (Figure~\ref{fig:pac}), we effectively applied SARIMAX to predict the number of taxi rides in Figure~\ref{fig:sari}. This approach allowed us to account for regular fluctuations over time and external factors influencing demand, providing distinct predictions and key findings into the temporal dynamics of taxi usage.

\subsection{Clustering}

\begin{figure}[h]
    \centering
    \includegraphics[width=1.0\textwidth]{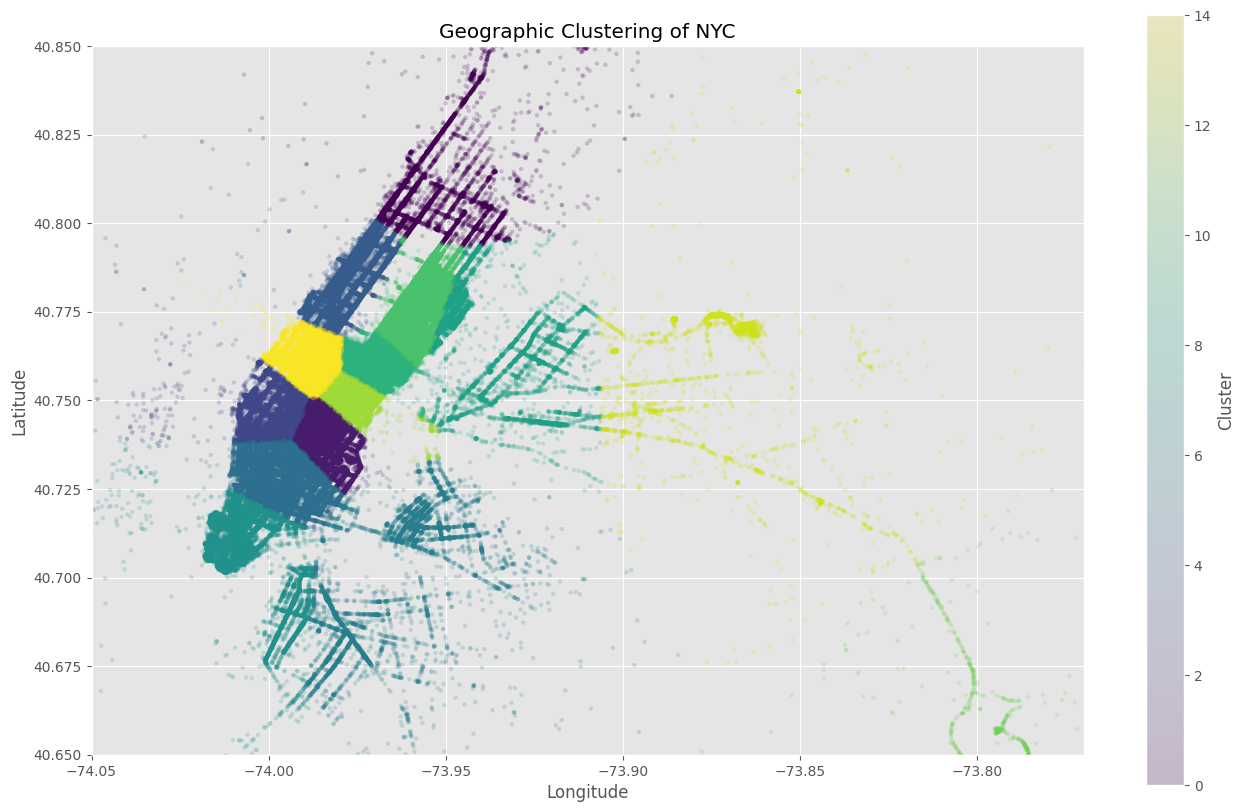}
    \caption{Geographical Clustering of NYC Taxi Trips}
    \label{fig:clus}
\end{figure}

For the clustering analysis, we applied K-means clustering with 15 clusters on the latitude and longitude coordinates of the pickup and drop-off locations in Figure~\ref{fig:clus}. This method involves grouping similar geographic locations into clusters, which helps identify patterns and significant regions within New York City. The clusters were visualized using a scatter plot, where each point represents a location and is colored according to its cluster assignment. This visualization effectively highlights areas with high or low demand for taxi services.

\section{Result Analysis}

\subsection{Descriptive Statistics}

The increase in pickups from the middle to the end of the month, as depicted in Figure~\ref{fig:image1}, indicates a period of heightened demand that operators can capitalize on by adjusting driver schedules and resource allocation. By aligning shifts with these peak demand periods, operators can improve service efficiency and maximize revenue opportunities. Similarly, the substantial rise in pickups during late hours (15th and 23rd hour), shown in Figure~\ref{fig:image2}, suggests that extending driver shifts into the late night could effectively address increased demand and reduce missed ride opportunities. Additionally, the findings that single-passenger trips are most common (Figure~\ref{fig:image3}) and that trip durations extend later in the day also tends to increase during the weekends (Saturday and Sunday)(Figures~\ref{fig:image6} and \ref{fig:image5}) offer further insights into optimizing fleet management and fare estimation. According to a New York Times article\footnote{\url{https://www.nytimes.com/2019/11/01/travel/weekend-trips.html}}, for most people in NYC, long weekends usually run from Friday to Sunday, a surge that often means higher travel costs, meaning the trip durations tend to increase during these weekends. Understanding these patterns allows taxi operators to anticipate periods of increased demand, particularly during long weekends, and adjust their fleet deployment and pricing strategies accordingly. This not only helps in better managing resources but also in maximizing revenue opportunities by aligning service availability with customer needs. Adjusting for longer trip durations on weekends and during evening hours can help in setting more accurate fares and planning efficient routes. By following these insights, taxi operators can better predict demand patterns and adjust their fleet accordingly. This ensures a more reliable service, leading to higher customer satisfaction. Ultimately, these strategic adjustments can significantly improve profitability.

From an insight by McKinsey and Company\footnote{\url{http://surl.li/qdkkct}}, Online Food orders spike on weekends as people prefer to order online more during weekends than weekdays. The analysis of the Pathao dataset shows that demand is highest on weekends (Friday and Saturday) as illustrated in Figure~\ref{fig:im1}. Figure~\ref{fig:ima3} reveals the top 10 most ordered items and cuisines. Through this analysis Food delivery owners can enhance their operations by increasing resources and staff on peak days (Fridays and Saturdays) to manage higher order volumes effectively. Furthermore, understanding which items and cuisines are most popular enables businesses to streamline their inventory and create focused promotions. This approach enhances customer satisfaction by more effectively meeting their preferences. In the long run, these strategies can increase sales and support business expansion.

\subsection{Geospatial Patterns}

The Geospatial analysis in Figure~\ref{fig:study} shows that Manhattan has the highest number of taxi pickups all day long, while Staten Island has the fewest. This means that Manhattan is a busy area where taxis are always in demand, whereas Staten Island sees less activity. Operators can use this analysis to focus their resources where they’re needed most. By placing more taxis in Manhattan, they can better serve the high demand and increase their earnings. In contrast, they might reduce the number of taxis in Staten Island or offer special promotions to boost business there. This helps in managing the fleet more effectively and improving overall profits.

Queens has the highest average trip duration and distance, with Staten Island having the lowest. Taxi operators can use this information to optimize fleet deployment. By ensuring more taxis are available in high-demand areas like Manhattan, they can better meet customer needs. Additionally, being prepared for longer trips in Queens can improve operational efficiency. These measures will ultimately enhance overall customer satisfaction. Additionally, operators can adjust pricing strategies based on trip distances and durations, maximizing revenue. Understanding these patterns can also help in planning driver shifts and routes to reduce idle time and enhance overall service.

\subsection{Temporal Trends}

For the time series analysis, we employed the SARIMAX model with parameters order=(1,0,1) and seasonal order=(1,0,1,7). The order parameters were selected based on model fitting diagnostics, combining autoregressive and moving average components that best fit the data. The seasonal order of (1,0,1,7) was chosen to account for weekly seasonality, given our dataset spans a month and thus a weekly cycle is a significant seasonal component. The analysis included evaluating the model's performance with RMSE metrics. The in-sample RMSE was 734.9441, and the out-of-sample forecast RMSE was 204.1525. This indicates that while the model performed reasonably well, there was a noticeable difference between in-sample and out-of-sample predictions. Finally, by examining the results, we can conclude that the SARIMAX model effectively captured the weekly seasonality with a RMSE of 734.944. This model can help operators by forecasting demand patterns, allowing for optimized fleet management and efficient driver scheduling. By predicting peak times and weekly trends, operators can adjust resources and pricing strategies to enhance service and increase revenue.

\subsection{Clustering Insights}
The K-means clustering analysis revealed distinct geographic patterns in taxi demand across New York City. By identifying clusters with high and low demand, taxi operators can strategically position their fleet in high-demand areas and allocate resources more effectively. This approach helps in optimizing service coverage, reducing wait times, and improving overall operational efficiency. 

\balance

\section{Discussion}

This study examined NYC Taxi and Pathao Food Delivery data to uncover key patterns in demand and service usage for urban transportation. Our analysis revealed that Manhattan consistently experiences the highest taxi pickup frequency throughout the day, reflecting its status as a major hub of activity. In contrast, Staten Island shows significantly lower pickup rates. For Pathao Food Delivery, demand peaks on weekends, particularly Friday and Saturday, with a clear preference for certain popular items and cuisines. The SARIMAX model effectively forecasted demand trends by capturing both seasonal and weekly variations, providing insights into fluctuations in taxi needs. The analysis showed that taxi trips take longer after midday. This suggests increased congestion or longer distances. For food deliveries, there are distinct patterns related to peak demand periods. These findings can guide operators in optimizing schedules, improving resource allocation, and enhancing customer satisfaction.

\section{Conclusion}
Our work provides a comprehensive analysis of transportation and delivery patterns in urban areas using the NYC Taxi and Pathao Food Delivery datasets. Our findings highlight critical trends such as peak times for taxi pickups and food delivery demand, which can assist operators in optimizing scheduling, resource allocation, and improving customer satisfaction. The use of time series analysis with SARIMAX and Geospatial clustering techniques offers valuable insights into both temporal and spatial dynamics. However, the study's scope was constrained by the resource, limiting our analysis to a single month (January) for the NYC Taxi Trip Dataset. This restriction may affect the broader applicability of the results and suggests the need for further research using more extensive datasets to validate and refine these insights.

\bibliographystyle{unsrt}  


\end{document}